\documentclass[%
 aip,
rsi,%
 amsmath,amssymb,
 reprint,%
]{revtex4-1}

\usepackage{graphicx}
\usepackage{dcolumn}
\usepackage{bm}
\usepackage{soul}
\usepackage{color}
\begin{document}
\newcommand{\ez}[1]{{\color{magenta}#1}}
\newcommand{\nic}[1]{{\color{cyan}#1}}
\newcommand{\jr}[1]{{\color{blue}#1}}
\preprint{AIP/123-QED}

\title{Rheological investigation of gels formed by competing interactions: a numerical study}

\author{Jos\'e Ruiz-Franco}
\email{jose.manuel.ruiz.franco@roma1.infn.it}
\affiliation{Dipartimento di Fisica, Sapienza Univesit\`a di Roma, Piazzale Aldo Moro 2, 00185 Roma, Italy}
\author{Nicoletta Gnan}%
\email{nicoletta.gnan@roma1.infn.it}
\affiliation{Dipartimento di Fisica, Sapienza Univesit\`a di Roma, Piazzale Aldo Moro 2, 00185 Roma, Italy}
\affiliation{CNR-ISC, UOS Sapienza, 00185 Roma, Italy
}%
\author{Emanuela Zaccarelli}
\email{emanuela.zaccarelli@cnr.it}
\affiliation{Dipartimento di Fisica, Sapienza Univesit\`a di Roma, Piazzale Aldo Moro 2, 00185 Roma, Italy}
\affiliation{CNR-ISC, UOS Sapienza, 00185 Roma, Italy
}%

\date{\today}

\begin{abstract}
A transition from solid-like to liquid-like behavior occurs when colloidal gels are subjected to a prolonged exposure to a steady shear. This phenomenon, which is characterized by a yielding point, is found to strongly depend on packing fraction.  However, it is not yet known how the effective inter-particle potential affects this transition. To this aim we present a numerical investigation of the rheology of equilibrium gels in which a short-range depletion is complemented by a long-range electrostatic interaction. 
We observe a single yielding event in the stress-strain curve, occurring at a fixed strain.
The stress overshoot is found to follow a power-law dependence on P\'eclet number, with an exponent larger than that found in depletion gels, suggesting that its value may depend systematically on the underlying colloid-colloid interactions.
We also establish a mapping between equilibrium states and steady states under shear, which allows us to identify the structural modifications induced by the presence of the shear. Remarkably, we find that steady states corresponding to the same P\'eclet number, obtained by different combinations of shear rate and solvent viscosity, show identical structural and rheological properties. Our results highlight the importance to understand the coupling between colloidal interactions, solvent effects and flow to be able to describe the microscopic organization of colloidal particles under shear.
\end{abstract}

\maketitle


\section{\label{sec:level0}Introduction}

Colloidal particles often form disordered arrested states, such as glasses and gels. Depending on the colloid-colloid interactions, different kinds of glasses can be found, including attractive, repulsive or Wigner glasses~\cite{pham2002multiple,zaccarelli2009colloidal,bonn1999aging,ruzicka2011fresh}. Similarly colloidal gels can be formed by different routes~\cite{gleim1998does}. An important distinction can be made to distinguish out-of-equilibrium gels formed via arrested spinodal decomposition from cases where gelation is obtained in equilibrium~\cite{sciortino2017equilibrium}. 
Non-equilibrium gels are found when colloids interact via hard-sphere-like excluded volume complemented by an isotropic short-range attraction, which is typical of depletion effects  induced by non-adsorbing polymer chains. At high enough depletion strength, a colloid rich-colloid poor phase separation takes place, in which the dense phase undergoes dynamical arrest into a gel state which interrupts the spinodal decomposition process~\cite{lu2008gelation}. On the other hand, gels can be obtained in equilibrium from a homogeneous fluid state, when attraction is not isotropic (e.g. patchy or limited valence\cite{sciortino2017equilibrium}) or when this is counter-acted by an additional long-range repulsion due to charge effects~\cite{groenewold2001anomalously,stradner2004equilibrium,sciortino2004equilibrium}. In this case, the competition between short-range attraction and long-range repulsion is able to avoid phase separation~\cite{campbell2005dynamical,sciortino2005onedimensional}.
Since colloidal gels are widely used for a variety of applications, including biomedical purposes~\cite{khetan2011patterning,guvendiren2012shear} food processing~\cite{mezzenga2005understanding}, optical sensing, thermoelectrics or catalysis~\cite{gaponik2011colloidal}, it is fundamental to control the gelation process and to be able to discriminate among the wide zoology of colloidal gels. 

To study the behavior of colloidal arrested states, one important experimental tool is rheology. In general, the application of a shear flow causes the occurrence of a solid-like to liquid-like transition that is preceded by a stress overshoot $\Sigma_{yield}$ at the yielding point. This indicates the maximum stress that the system can accumulate~\cite{liddel1996yield,mason1996yielding}. After yielding, the system is able to approach a steady state with liquid-like behavior. The way in which this steady state is reached of course depends on the studied system and on the shear conditions. However, different types of arrested states generally respond in a different manner. For the widely studied hard-sphere glasses, for which excluded volume interactions are responsible for the kinetic arrest, a single yield mechanism is observed~\cite{derec2003aging,petekidis2004yielding}. This is normally attributed to cage breaking only or to particle exchange with their nearest neighbours. On the other hand, for attractive glasses, which can be induced by depletion interactions at relatively high packing fractions $\phi$, two different yielding points have been reported~\cite{pham2004glasses,pham2008yielding}. A first one ($\Sigma_{yield 1}$) is associated to bond breaking at local level with the system retaining a solid-like character, while a second one ($\Sigma_{yield 2}$) is related to a structural rearrangement (cage-breaking) after which the system is able to flow. Between the two yielding events, there is a local bond reorganization which, however, does not significantly alter the system. A double yielding mechanism has also been observed for depletion-induced gel-like samples at $\phi\sim 0.40$, \cite{koumakis2011two} an effect that could possibly be associated to the relatively large packing fraction. Indeed, the caging effect disappears for $\phi\lesssim0.2$,\cite{koumakis2011two} leaving only one yielding point associated to bond-breaking for low density gels. In addition, it is interesting to note that for very large attraction strengths $\left(\sim100k_{B}T\right)$, experiments on very dilute depletion gels also reported the occurrence of two yielding mechanisms~\cite{chan2012two}: in this case the bond breaking yielding point observed for weaker attractions is preceded by another yielding point associated to the onset of bond rotation.
Furthermore, it is important to consider that the presence of an imposed shear flow naturally induces anisotropy or heterogeneity in the structure of the system~\cite{varga2018large}, which affects the rheological properties~\cite{vermant2005flow}.  This aspect has been tackled both  in experiments~\cite{rajaram2010microstructural,min2014microstructure,eberle2014shear} and simulations~\cite{colombo2014stress,park2017structure,moghimi2017residual,boromand2017structural,jamali2017microstructural,johnson2018yield}. 

To our knowledge, no rheological studies ---either numerical or experimental---  have been performed to date on gels obtained via the competition of short-range attraction and long-range repulsion. The aim of the present work is to fill this gap by investigating this type of gels via numerical simulations performing start-up shear experiments for a wide variety of steady shear conditions. Mainly, we focus on Langevin Dynamics (LD) simulations, which is appropriate to treat colloids in an implicit solvent. With this approach, we provide evidence that, while some rheological features of depletion-induced gels are also found in our case, others reveal that the inter-particle potential plays an important role. Indeed the long-range repulsion makes our gels more resistant to shear flow, with a modified dependence of the stress overshoot with respect to shear (quantified by P\'eclet number). Similarly, we study the implication of the presence of shear flow at the microscopic level, focusing on the evolution of network bonds and the anisotropy. Since the competition between Brownian motion and shear flow modifies the gel response, we show that a different balance of the two effects allows to produce different anisotropies in the sheared system. Interestingly, we find that the resulting anisotropic patterns take a characteristic form when the gel undergoes crystallization under shear. As a next step we focus on the role of solvent on mechanical response and microscopic gel restructuring. To this end, we also implement Molecular Dynamics (MD) simulations, where the solvent is neglected. This has the aim to clarify the influence of the microscopic dynamics on the sheared systems, a practice that has been carried out for example in glassy systems, where it was found that the long-time dynamics does not depend on the presence of a solvent\cite{gleim1998does}.
Connecting equilibrium and steady states via a mapping at equal potential energy allows us to highlight the effect of the shear in counter-acting the long-range repulsion, while not significantly altering the local (average) structure as compared to gels without shear. We find that steady states obtained at equal P\'eclet number are identical to each other, independently of the effective solvent viscosity used in the simulations. However, the comparison with MD simulations reveals that a different microscopic dynamics inevitably alters the rheological response of the gel, but does not affect its average thermodynamic properties. To validate the robustness of our results, we also repeat some simulations for larger system sizes finding that no size effects are observed for the steady state properties that we have considered. Some differences however arise in the transient immediately after the start-up of the shear. 

The paper is organized as follows. In Section~\ref{sec:level1} we describe the numerical simulations for the different methods that we use and we also define all the observables that are calculated in this study. In Sec.~\ref{sec:level2}, we present our results discussing the microscopic organization of the gel in the presence of shear (Sec.~\ref{sec:level3}), the anisotropy generated by the competition between Brownian dynamics and shear flow (Sec.~\ref{sec:level4}), and the shear-induced crystalline structures formed by the gel (Sec.~\ref{sec:level4bis}). We also show the results of the mapping between equilibrium states and stable states under shear (Secs.~\ref{sec:mapping}-\ref{sec:level5}) and we conclude the section by addressing the problem of system size effects (Sec.~\ref{sec:level7}). Finally, in Sec.~\ref{sec:level8}, we discuss our results and present the conclusions of our study.

\section{\label{sec:level1}Methods: Simulations and theory}
\subsection{Simulation details}
\begin{figure}
\includegraphics[width=0.75\linewidth]{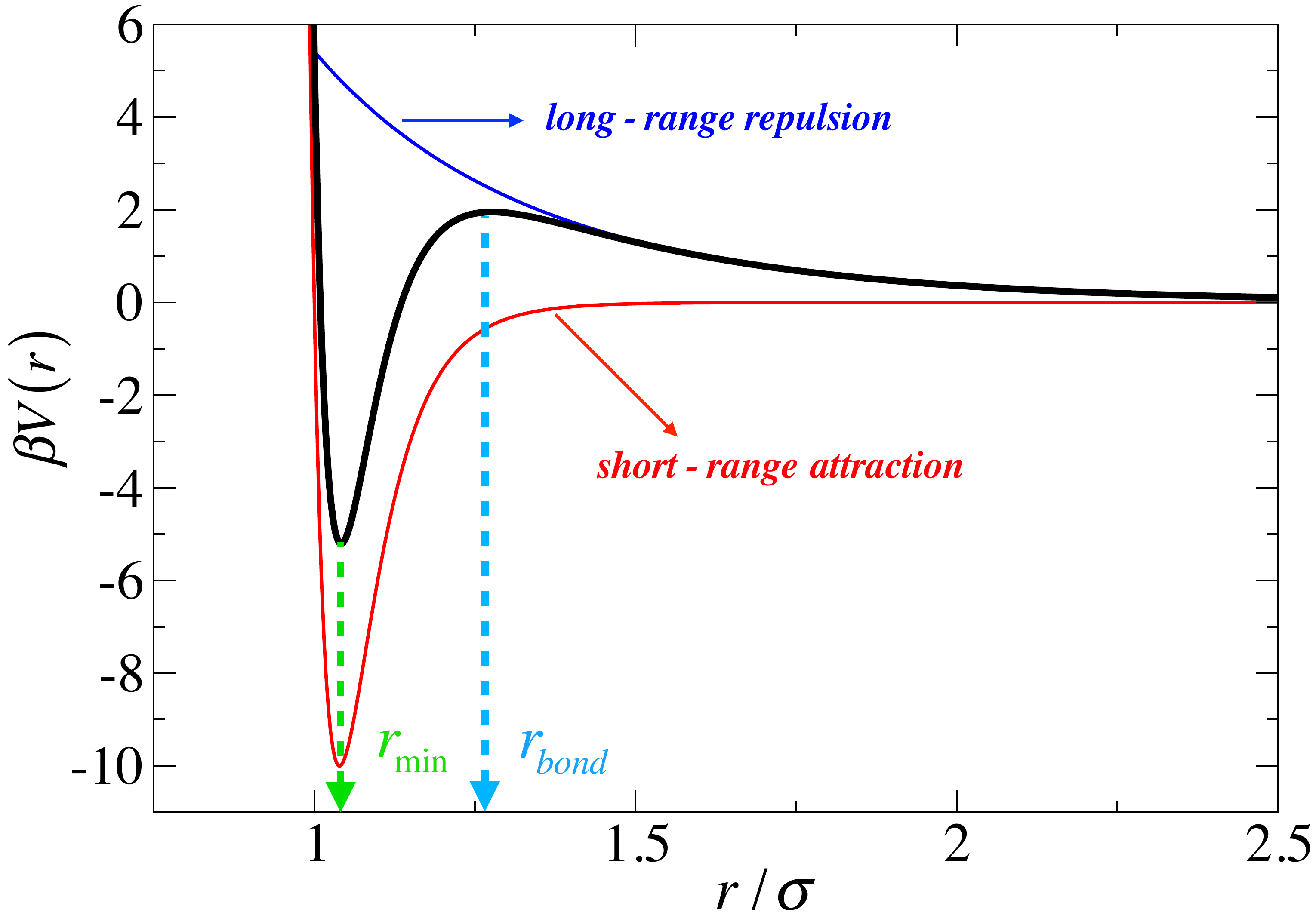}
\caption{Inter-particle potential (black curve) given by the sum of a short-range depletion attraction (red curve) and a long-range electrostatic (blue curve) repulsion. The arrows indicates the position of the global minimum of the potential $r_{min}$ and of the bond distance $r_{bond}$.}
\label{fig:Potential}
\end{figure}

We perform simulations of monodisperse colloids of diameter $\sigma$ 
and mass $m$ in presence of steady shear. Most of the simulations are carried out with $N=2000$ colloidal particles, but in order to assess size effects onto the shear results, we also repeat some simulations for $N=5000$ and $N=10000$ colloids.
Particles interact via a potential $V(r)$ which is the sum of a short-range attraction and a long-range repulsion, \cite{sciortino2004equilibrium,sciortino2005onedimensional} as
\begin{equation}
	\label{eq:SR}
	V(r)=4\epsilon\left[\left(\frac{\sigma}{r}\right)^{2\alpha}-\left(\frac{\sigma}{r}\right)^{\alpha}\right]+A \frac{e^{-\kappa r}}{r/\sigma}.
\end{equation}
Here the short-range attraction (mimicking depletion interactions) is modeled as a generalised Lennard-Jones potential \cite{vliegenthart1999strong} with the potential depth $\epsilon$ and the particle diameter $\sigma$ being, respectively, the units of energy and length, while the long-range repulsion (representing a screened electrostatic contribution) is described by a Yukawa potential where $\kappa$ is the inverse of the Debye screening length and $A$ is the repulsion amplitude. Time is measured in units of $\sqrt{(m\sigma^2)/\epsilon}$.
Following Ref. \cite{sciortino2005onedimensional}, we fix $\alpha=18$, $A=4\epsilon$ and $\kappa=2\sigma^{-1}$. The resulting interaction potential is illustrated in Fig.~\ref{fig:Potential}. In our simulations we fix $k_{B}=1$ and use a cutoff of the interactions at  $r_{c}=4\sigma$. Particles interacting with this potential are able to form an equilibrium gel at low/intermediate packing fractions $\phi$ and sufficiently low temperatures \cite{sciortino2005onedimensional,zaccarelli2007colloidal}. In this work we consider a packing fraction $\phi=\frac{\pi}{6}\sigma^{3}\frac{N}{V}$=0.16 and $T=0.1$, where $V$ is the volume of the cubic simulation box. To reduce statistical noise, data are always averaged over three independent realizations. 

In our study we first prepare a gel state in equilibrium and then we perform a start-up shear test by applying a steady shear flow onto the gel imposing the so-called Lees-Edwards boundary conditions~\cite{lees1972computer}. We consider the gradient velocity to be in the $\hat{z}$ direction while the shear velocity is in the $\hat{x}$ direction, so that the shear rate is defined as $\dot{\gamma}\equiv v_{x}/z$ and is measured in inverse time units. We employ a Langevin thermostat which acts on the so-called peculiar velocity\cite{shang2017assessing,ruiz2018effect}, which is defined for particle $i$ as $v'_{i,x}=v_{i,x}-u_x(z)$ where $v_{i,x}$ is the $x$-component of the particle thermal velocity and $u_x(z)=\dot\gamma z$ is the stream velocity. Simulations are performed in the canonical ensemble $NVT$ and the properties of the solvent are controlled by its viscosity $\eta$, which in turn determines the friction coefficient $\xi=3\pi\eta \sigma$.

To monitor how the microscopic dynamics affects the shear response of the gel, we also perform simulations in the absence of an implicit solvent. To this aim, we use a so-called Gaussian thermostat, for which the temperature is controlled by imposing a constant kinetic energy (iso-kinetic ensemble)~\cite{morriss1998thermostats}. The equations of motion are then solved by means of the SLLOD integrator~\cite{zhang1999kinetic}.

\subsection{Definition of a generalized P\'eclet number} 
The strength of the applied shear is usually quantified by the P\'eclet number $Pe$, which controls the balance between Brownian motion and shear effects: the behaviour of the system is essentially governed by Brownian motion for   $Pe <$ 1, while  it is dominated by the shear flow for $Pe >$1.  However, in its standard definition\cite{cloitre2010high}, the P\'eclet number is taken to be equal to $\dot\gamma\tau_{B}$,  where $\tau_{B}=\frac{\sigma^{2}}{4D}$ is the the Brownian time and $D=\frac{k_{B}T}{\xi}$ is the diffusion coefficient of the particle at infinite dilution. This definition is not appropriate to quantify the strength of the shear on the gel state because the system is far from dilute conditions and cannot be simply generalized because the diffusion coefficient in the gel tends to zero. To provide a meaningful definition of $Pe$, we adopt the modified definition of Ref.\cite{koumakis2011two} 

\begin{equation} 
	\label{eq:Pegel}
	Pe=\frac{F_{visc}}{F_{bond}}
\end{equation}

\noindent
which quantifies the resistance of the bonds between two colloids to the shear. Here $F_{visc}$  is the drag force that is able to displace two particles up to a distance larger than the attractive range of the potential:

\begin{equation} 
	\label{eq:Fvisc}
	F_{visc}=\xi v_{s} 
\end{equation}

\noindent
where $v_s=\dot{\gamma}r_{min}$ and $r_{min}\approx1.05\sigma$ is the global minimum of potential, i.e. the equilibrium distance of two particles (see Fig.~\ref{fig:Potential}). Instead, $F_{bond}$ is the bonding force which is responsible for maintaining the bond between the two colloids. By increasing the interparticle distance $F_{bond}$ will increase until the two particles become more far than a maximum distance $r_{bond}$, which is the maximum of the potential, after which the effective force becomes repulsive. In the present case, $r_{bond}=1.28\sigma$ as shown in Fig.~\ref{fig:Potential}. We thus define the bonding force as the variation of energy in the attractive range of the effective potential $\Delta r=r_{bond}-r_{min}$, as

\begin{equation} 
	\label{eq:Fbond}
	F_{bond}=\frac{\Delta V\left(r\right)}{\Delta r}=\frac{V\left(r_{bond}\right)-V\left(r_{min}\right)}{r_{bond}-r_{min}}.
\end{equation}

\noindent
Using the definition of eq.(\ref{eq:Pegel}) we find that when the two forces are balanced, i.e. $Pe=1$, the original P\'eclet number calculated under dilute conditions would be much higher, i.e. $\approx320$.

We notice that the P\'eclet number can be defined only for simulations in the presence of a solvent such as LD ones. For MD simulations we thus quantify the strength of the shear by varying the shear rate only.

\subsection{Calculated observables}
During application of the steady shear, we calculate the internal stress tensor $\Sigma_{xz}$ using the Irving-Kirkwood expression~\cite{irving1950statistical}:
\begin{equation} 
	\label{eq:Stress_Tensor}
	\Sigma_{xz} = \frac{1}{V}\left\langle\sum_{i}[m_{i}v'_{i,x}v_{i,z}+\sum_{j>i}r_{ij,x}F_{ij,z}]\right\rangle
\end{equation}
where $r_{ij}$ and $F_{ij}$ are, respectively, the distance and the force between particles $i,j$ and the brackets $\langle \ldots \rangle$ represent the ensemble average.

To provide a microscopic understanding of the stress tensor behavior, we monitor the time evolution of the bond organization between the particles,  calculating: (i) the fraction of bonds $f_b$ defined as the number of bonds in a given configuration under shear divided by the number of bonds that were present in the system  prior to switching on the shear; (ii) the fraction of unbroken bonds $f_u$ that were also present at zero-shear; (iii) the bond potential energy $E_{b}$~\cite{landrum2016delayed}, defined as the absolute value of the average potential energy between all pairs of bonded particles, normalized to its initial value, again prior to switching on the shear flow. 

Next, we examine the changes in the structure by calculating the static structure factor: in equilibrium this is defined as 
\begin{equation} 
	\label{eq:Sq_eq}
	S_{eq}(q)=\frac{1}{N}\left\langle \sum_{ij}e^{-i\mathbf{q}\left(\mathbf{r}_{i}-\mathbf{r}_{j}\right)}\right\rangle,
	\end{equation}
while when shear is applied we evaluate it in  the velocity-vorticity plane at $q_{z}=0$, as 
\begin{equation} 
	\label{eq:Sq_shear}
	S_{shear}\left(q_{x},q_{y},0\right)=\frac{1}{N}\left\langle \sum_{ij}e^{-i[\mathbf{q_{x}}\left(x_{i}-x_{j}\right)+\mathbf{q_{y}}\left(y_{i}-y_{j}\right)]}\right\rangle. 
	\end{equation}
For simplicity, we refer to both types of structure factors as $S(q)$ in the main text, since they depend only on the modulus of the wavevector, being it calculated in 3D or in 2D along the velocity-vorticity plane. 

A complementary picture of the structure, which allows to identify the anisotropy induced by shear, is provided by using a suitable expansion of the pair correlation function $g\left(\mathbf{r}\right)$. In particular, we consider the expansion 
\begin{equation} 
g\left(\mathbf{r}\right)=g_{s}\left(r\right)+\sum_{l=1}^{\infty}\sum_{m=-l}^{l}g_{l}^{m}\left(r\right)Y_{l}^{m}\left(\theta\phi\right)
\end{equation}
\noindent  into spherical harmonics $Y_{lm}\left(\theta,\phi\right)$~\cite{hess1983pressure,hanley1987shear}, where $g_{s}\left(r\right)$ is the usual (isotropic) radial distribution function~\cite{hansenbook} and the expansion coefficients are $g_{lm}\left(r\right)=\int g\left(\mathbf{r}\right)Y_{lm}^{*}d\Omega$ with $\Omega$ the solid angle and $d\Omega=sin\theta d\theta d\phi$. Our colloidal gel is made up of identical particles, implying that $g\left(\mathbf{r}\right)=g\left(-\mathbf{r}\right)$. This ensures that only even values of $l\geq2$ have to be accounted for~\cite{hess1983pressure,hanley1987shear}. However, coefficients with $l>2$ are of small amplitude, so that, in general, it is sufficient to consider only the term with $l=2$~\cite{zausch2009dynamics}. In addition, due to the geometry of the planar Couette flow,  only coefficients with $m=\pm 2$ are in the shear flow plane and the only non-zero contribution comes from their imaginary part. We thus focus on the imaginary part with $m=-2$, i.e. $Im\:g_{2}^{-2}\left(r\right)$, which is calculated as~\cite{zausch2009build}:

\begin{eqnarray}
	\label{eq:Gr_an}
	&&Im \:g_{2}^{-2}\left(r\right) = \sqrt{\frac{15}{8\pi}}\frac{L^{3}}{N^{2}} \times \nonumber \\
	&&  \left \langle \sum_{i}^{N}\sum_{j\ne i}^{N}\delta\left(\left| \mathbf{r}_{i}-\mathbf{r}_{j} \right|-r\right)\frac{\left(x_{i}-x_{j}\right) \left(z_{i}-z_{j}\right)}{r^4}\right\rangle.
\end{eqnarray}

\noindent This function is characterized by the presence of two peaks: the first one is a minimum signalling the accumulation of the particles along the compression axis, while the second one is a maximum which corresponds to the depletion of the particles along the extension axis~\cite{park2017structure,khabaz2017shear}.


To identify solid-like particles we use the local bond-order analysis introduced by Steinhardt et {\it al.}~\cite{steinhardt1983bond}, where the complex vector $q_{lm}\left(i\right)$ of particle $i$ is defined as $q_{lm}\left(i\right)=\frac{1}{N_{b}\left(i\right)}\sum_{j=1}^{N_{b_{i}}}Y_{lm}\left(\hat{r}_{ij}\right)$, where $N_{b}\left(i\right)$ is  the set of bonded neighbours of particle $i$ and $\hat{r}_{ij}$ is the unit vector specifying the orientation of the bond between particles $i$ and $j$. Using the complex vectors $q_{6m}$, we are able to assign a solid connection between particles $i$ and $j$ if $d_{6}\left(i,j\right)=\sum_{m=-6}^{6}q_{6,m}\left(i\right)\cdot q{}_{6,m}^{*}\left(j\right) \geq 0.7$. A particle is defined to be solid-like if it has 6 or more solid connections with its neighbours~\cite{pusey2009hard}. The percentage of solid particles is thus defined as $ X\left(t\right)=\frac{N_{X}}{N}$ with $N_{X}$ the number of solid-like particles. Following 
Ref. [60], we also calculate the rotationally invariant bond order parameters $\overline{q}_{l}\left(i\right)$ and $\overline{w}_{l}\left(i\right)$. To define these two parameters, it is necessary to compute the averaged local bond order parameters:

\begin{equation}
\label{eq:qlm}
	\overline{q}_{lm}\left(i\right)=\frac{1}{\widetilde{N}_{b}\left(i\right)}\sum_{j=0}^{\widetilde{N}_{b}\left(i\right)}q_{lm}\left(j\right)
\end{equation}

\noindent where $\widetilde{N}_{b}\left(i\right)$ is the number of neighbours including the particle $i$ itself. In this way, the first invariant bond order parameter is defined as

\begin{equation}
\label{eq:qlm}
	\overline{q}_{l}\left(i\right)=\sqrt{ \frac{4\pi}{2l+1}\sum_{m=-l}^{l} \left|\bar{q}_{lm}\left(i\right)\right|^{2} }
\end{equation}

\noindent while the second one is defined as

\begin{equation}
\label{eq:wlm}
\begin{split}
	\overline{w}_{l}\left(i\right) = \frac{\underset{m_{1}+m_{2}+m_{3}=0}{\sum} \left( \begin{array}{ccc}
l & l & l\\
m_{1} & m_{2} & m_{3}
\end{array} \right) \overline{q}_{lm_{1}}\left(i\right)\overline{q}_{lm_{2}}\left(i\right)\overline{q}_{lm_{3}}\left(i\right)}{\left( \sum_{m=-l}^{l} \left| \overline{q}_{lm}\left(i\right) \right|^{2} \right)^{3/2}}
\end{split}
\end{equation}

\noindent where the term in parentheses is the Wigner 3-j symbol. The integers $m_{1}$, $m_{2}$ and $m_{3}$ runs from $-l$ to $l$ but only the combination that meets the requirement $m_{1}+m_{2}+m_{3}=0$. Using $l=4$ and $l=6$ it is possible to establish a separation between BCC, FCC and HCP structures \cite{russo2012microscopic}.

\section{\label{sec:level2}Results}


\subsection{\label{sec:level3}Microscopic organization of the gels under shear}
\begin{figure}
\includegraphics[width=\linewidth]{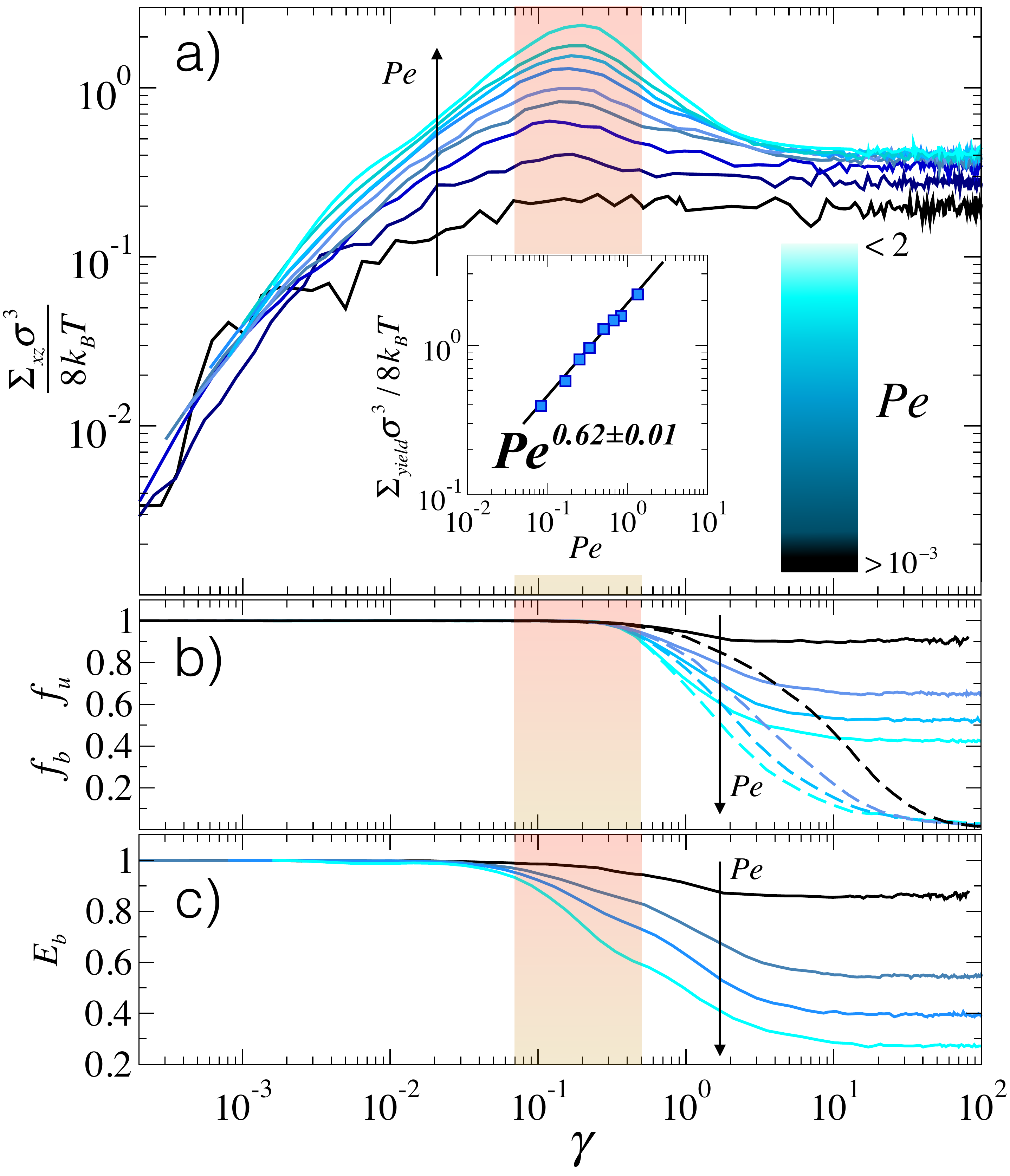}
\caption{\label{fig:Stress} Gel response to shear within LD simulations with $\xi=10^2$ for different $Pe$: (a) Normalized stress $\Sigma_{xz}a^{3}/k_{B}T$ versus strain $\gamma$. {\it Inset}: stress overshoot $\Sigma_{yield}$ as a function of $Pe$. The black line is a power-law fit to the numerical data (symbols); Strain dependence of (b) Fraction of bonds $f_{b}$ (solid lines), fraction of unbroken bonds $f_{u}$ (dashed lines) and (c) bond energy $E_{b}$, normalized to their values in the absence of shear. The shaded areas indicate the strain region in which yielding takes place.} 
\end{figure}

\begin{figure*}
\includegraphics[width=0.7\linewidth]{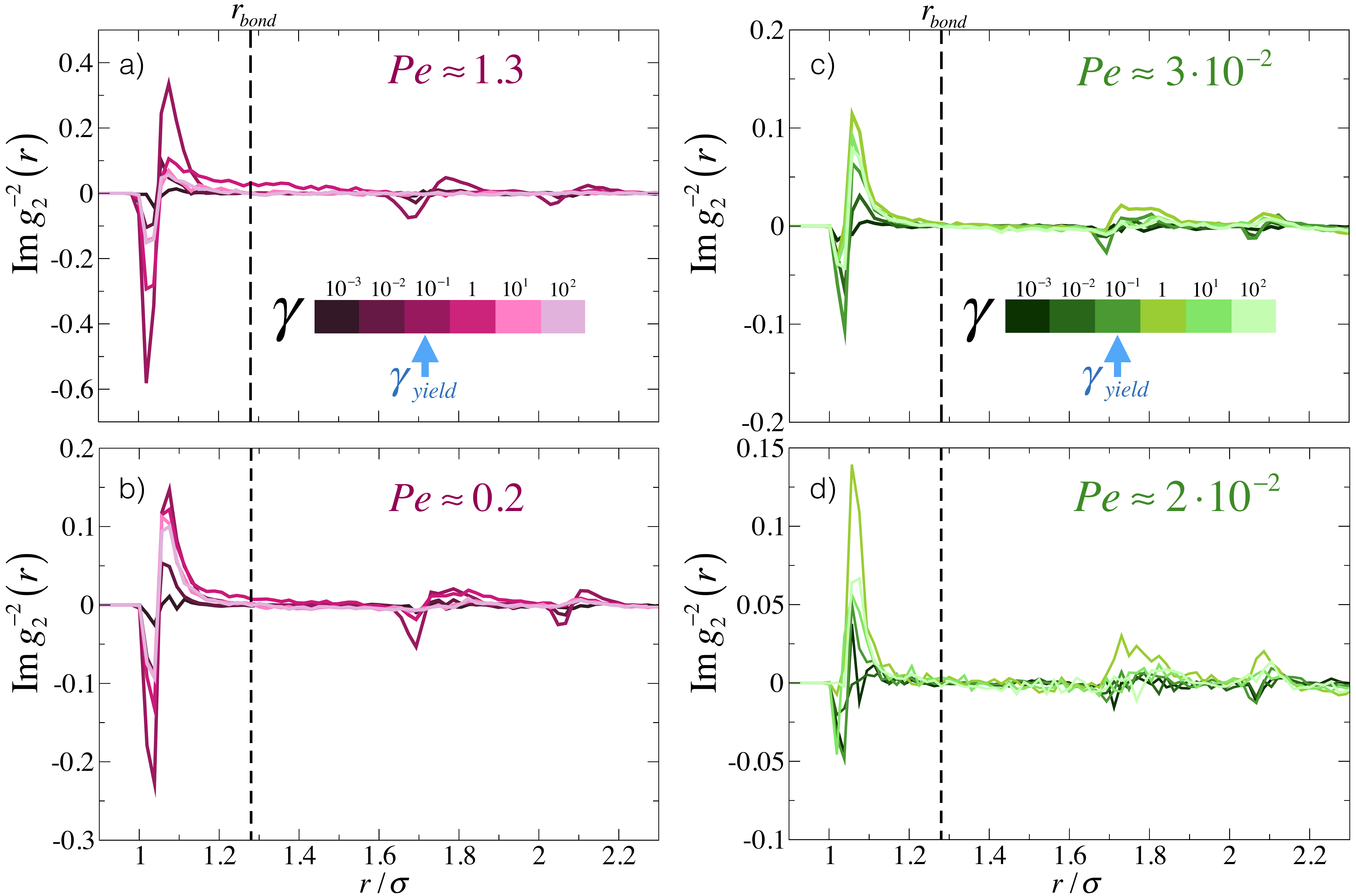}
\caption{Shear-induced anisotropy Im$g_{2}^{-2}\left(r\right)$ for LD simulations with $\xi=10^{2}$ at different (a) $Pe\approx1.3$ and (b) $Pe\approx0.2$ and with $\xi=10$ at different (c) $Pe\approx3\cdot10^{-2}$ and (d) $Pe\approx2\cdot10^{-2}$.}
\label{fig:Anisotropy}
\end{figure*}

We start by showing the behavior of the stress tensor under shear. To facilitate a comparison with experimental results on colloidal gels, we focus on the case of LD simulations, which take into account the presence of the solvent in an effective way, although neglecting hydrodynamic interactions. As discussed above, for several attractive arrested states~\cite{pham2008yielding,koumakis2011two,chan2012two} two yielding mechanisms have been observed. It is now interesting to see what happens in the present case of equilibrium gels obtained by competing interactions. The behaviour of the normalized stress tensor $\Sigma_{xz}\sigma^{3}/8k_BT$ is reported as a function of strain $\gamma=\dot{\gamma}t$  for several values of $Pe$. We find that only one yielding mechanism takes place in our gel, confirming the results reported for depletion-induced gels at comparable packing fraction and attraction strength~\cite{koumakis2011two}. Similarly to previous findings, the position of maximum of the stress tensor, defining the yielding point $\Sigma_{yield}$, is insensitive to shear rate~\cite{whittle1997stress,koumakis2011two}. We also find a power law dependence of the yield stress on the shear rate, i.e., $\Sigma_{yield}\propto$ Pe$^{\delta}$.  However, for depletion-induced gels at intermediate $\phi$ the power law exponent has been found to be $\delta\sim 0.5$ both in simulations~\cite{whittle1997stress} and experiments~\cite{koumakis2011two}. On the other hand, numerical simulations for the Derjaquin-Landau-Verwey-Overbeek (DLVO) potential~\cite{park2013structural} have found $\delta\sim 0.56$. We find that for our model the yield stress follows a power law with an exponent $\delta\sim 0.62\pm 0.01$ (see inset in Fig.~\ref{fig:Stress}(a)). This result suggests that there is a systematic change of $\delta$ with the employed interaction potential between the colloids.  Interestingly for dense colloidal glasses, an almost constant value of the stress overshoot with $Pe$ was found\cite{laurati2012transient}. To provide an interpretation of these findings, we refer to early theoretical studies on polymer~\cite{groot1995dynamic} and colloidal gels~\cite{whittle1997stress}, which reported a link between  $\delta$ and the fractal dimension $d_{f}$ of the system, i.e. $\delta=\frac{3-d_{f}}{2}$. Using this relationship, the fractal dimension would decrease from $\sim2.0$ to $\sim1.75$ as we add the long-range electrostatic repulsion to the short-range depletion one. This is in agreement with direct estimates of the fractal dimensions from assessment of the clusters in equilibrium~\cite{sciortino2005onedimensional}, where a very low value of $d_f\sim 1.25$ was found. It is plausible that this value is slightly increased by the presence of the shear. Thus, from this type of result we can get an indirect estimate of the gel structure from rheological measurements. It would be interesting to test this relationship to systems with different fractal dimensions.
We may also speculate that an increase of $\delta$ can  be interpreted in terms of a stronger resistance to the flow of the gel under shear at structural level. This interpretation is confirmed by  comparing our results with those presented in Ref.~\cite{johnson2018yield}, where numerical simulations of a gel were performed using a Morse potential with a minimum value of the energy $\sim 6k_{B}T$, similarly to our case, reporting an exponent $\delta \leq 0.5$. In our model, the presence of the long-range repulsion could possibly counteract the effect of the shear in breaking bonds, inducing an increase of $F_{bond}$ and thus, making our gel more resistant to shear flow, which is thereby reflected by an increase of $\delta$.

To deepen our microscopic understanding of the yielding point, we monitor the fraction of bonds $f_b$ and the bond energy $E_b$ in Fig.~\ref{fig:Stress} (b) and (c) for the same $Pe$ values. We find that at yielding the total number of bonds does not change significantly, while the bond energy already starts to decrease. Interestingly, if we separately monitor only the fraction of unbroken bonds $f_u$ with respect to the zero-shear case, we find no broken bonds until yielding. Thus, the gel network remains essentially unaffected up to $\Sigma_{yield}$ in agreement with Refs.~\cite{koumakis2011two,laurati2011nonlinear,johnson2018yield}.
The decrease of the energy however might indicate that bonds become more and more stretched under the shear deformation, prior to eventual breakage. After the yielding point, when the bond energy has already dropped a significant amount, $f_u$ rapidly decreases until reaching a steady state where basically none of the initial bonds is left intact. However, new bonds appear and indeed, $f_b$ reaches a new plateau in the steady state. In all cases, the system is found to form a new percolating network, whose structure is very different from the initial gel state as we will show in section~\ref{sec:mapping}.

\subsection{\label{sec:level4}Anisotropy}

From the microscopic point of view, the anisotropy induced by the shear flow has been studied to understand how the balance between Brownian motion and shear flow can affect the microscopic structure for different systems. We have thus studied the evolution of the anisotropy of our gels by calculating Im$g_2^{-2}(r)$, for different {\it Pe} (changing both $\dot{\gamma}$ and $\xi$) in the LD simulations. We incidentally notice that for hard sphere systems studied within Brownian dynamics an inversion of the peak amplitudes in Im$g_2^{-2}(r)$[60] was observed upon increasing $Pe$, while for low shear contribution, the compressional peak is more pronounced than the extensional one, this situation was found to be reversed at high shear. This was interpreted as a consequence of the net contribution of the shear with respect to the Brownian dynamics. At high $Pe$, the convective motion of hard spheres increases and hence a larger anisotropy is found in the extensional axis. For the current system, anisotropy results are reported in Figs.~\ref{fig:Anisotropy} (a,b) for $\xi=10^{2}$ and in Figs.~\ref{fig:Anisotropy} (c,d) for $\xi=10$. For each solvent condition, we have monitored the evolution of anisotropy in the system and associate it to the analysis of the stress and bonds reported above, including the yielding manifestation up to the steady state.

For $\xi=10^{2}$ we find that the anisotropy starts to grow already well before the yielding point (occurring for $\gamma\sim 10^{-1}$). A characteristic two-peaked shape is observed, which is made of a negative peak followed, at larger distances, by a positive one. These two peaks indicates the increase of anisotropy along the compressional and extensional axis, respectively~\cite{park2017structure}. Interestingly, the distance at which Im$g_{2}^{-2}\left(r\right)$ passes through zero corresponds to the minimum $r_{min}$ of the total interaction potential. In addition, we observe that the positive peak does not exceed the maximum bond distance $r_{bond}$, showing that this modification of the gel structure affects the bonds between the particles and not larger distances. At yielding, where still the network is intact but the maximum accumulation of stress occurs, the induced anisotropy is also maximum. It is found that anisotropy at this point is able to propagate at distances larger than the bond ones, affecting the whole structure, as shown by additional peaks arising for $r>r_{bond}$. When the bonds finally start to break, after yielding, the anisotropy also decreases and the peaks at larger distances disappear. Interestingly, after this happens, Im$g_{2}^{-2}\left(r\right)$ shows a long-range tail which is most evident in Fig.~\ref{fig:Anisotropy} (a,b). This tail could indicate that the dissipation of the anisotropy does not happen instantaneously but occurs within a finite time, in correspondence with the smooth decay of $f_{b}$ and $f_{u}$ after the yielding point. Indeed, at larger strains, the tail disappears. However, even in the steady state ($\gamma \gtrsim 10$), a significant amount of anisotropy still remains at local level. The reported behavior is enhanced for higher values of $Pe$, which in general induce a larger amount of anisotropy in the system~\cite{johnson2018yield}. The behavior described for hard-spheres systems above is not observed~\cite{foss2000structure,mewis2012colloidal}, probably due to the fact that our short-range repulsion, although very steep, is not hard-sphere-like, thus allowing for a moderate compression of the particles themselves. This can be seen in the the compressional peak moving at slightly smaller distances with increasing $Pe$. Indeed, a similar feature was observed in the anisotropy of deformable particles~\cite{khabaz2017shear}.

On the other hand, we find remarkable differences in the behavior of Im$g_2^{-2}(r)$ for $Pe\lesssim3\cdot10^{-2}$, as shown in Fig.~\ref{fig:Anisotropy}(c,d). In this regime, the Brownian motion is much stronger than the shear flow. We monitor this behavior for the lower studied friction coefficient ($\xi=10$) and find that, after the yielding point,  the extensional peak becomes much more pronounced while the compressional one tends to disappear. To clarify the effects due to the competition between Brownian dynamics and the shear flow, we compare the anisotropy distribution for two systems having the same shear rate $\dot{\gamma}=0.05$ but different friction coefficients: respectively $\xi=10^{2}$ in Fig.~\ref{fig:Anisotropy}(b) and $\xi=10$ in Fig.~\ref{fig:Anisotropy}(d). 
It is clear that, tuning the solvent properties at the same shear rate, we can manipulate the anisotropy induced in the system by the shear flow. 
While at high $Pe$ both extension and compression of the bonds take place, for low $Pe$ only the extensional axis grows. This feature, coupled to the lack of a clear yielding point (see Fig.~\ref{fig:Stress}), indicates that the system is undergoing a restructuring process. Under these conditions, the small perturbation acted by the shear flow onto the system is enough to allow the particles to reorganize towards a more ordered configuration. Indeed, a fluid-to-crystal transition is observed at higher strains, as shown in the next section. On the other hand, at large frictions (and hence higher $Pe$) the Brownian motion dominates the effect of the shear and crystallization does not occur, at least on the simulated time scales. Interestingly, a recent study of a jammed suspension under steady shear \cite{khabaz2017shear} reported a higher accumulation of particles along the compression axes prior to crystallization, an opposite result to the present case. However, in that system, particles interact with a soft Hertzian repulsion and thus, the high deformability should be responsible for the observed behavior. Despite the differences, this study shows that such a large asymmetry of the two peaks of Im$g_{2}^{-2}\left(r\right)$ is a distinctive signature of an incipient crystallization. It would be interesting to confirm this feature in other works, either numerical or experimental ones.

\begin{figure}
\includegraphics[width=\linewidth]{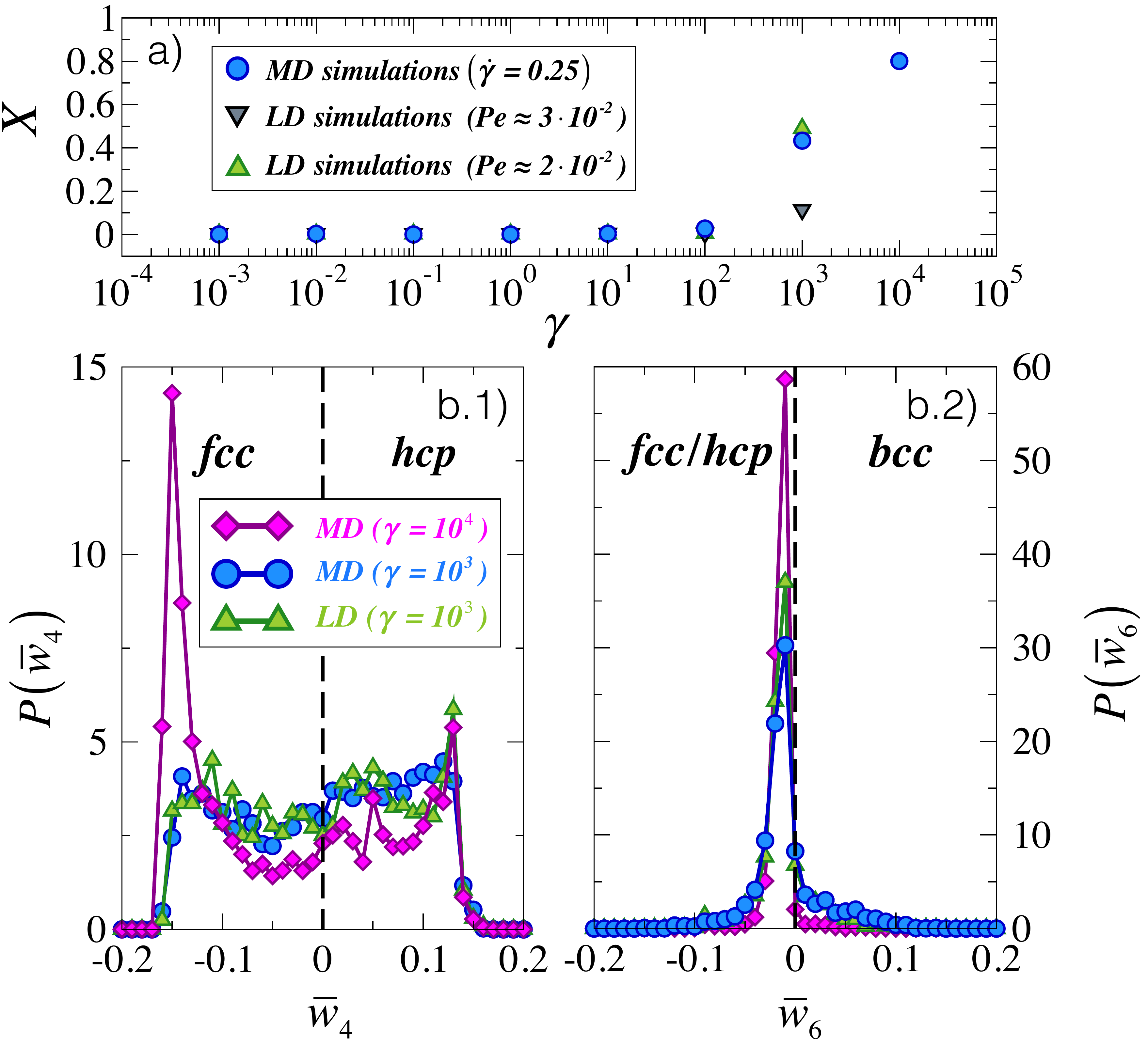}
\caption{\label{fig:Crystallization} (a) Number of solid-like particles $X$ versus strain for LD simulations with $\xi=10$ at $Pe\approx 3\cdot10^{-2}$ and $Pe\approx 2\cdot10^{-2}$ and MD simulations with $\dot{\gamma}=0.25$; (b) rotationally invariant bond order parameter distribution $P\left(\bar{w}_{4}\right)$ (b.1) and $P\left(\bar{w}_{6}\right)$ (b.2) when the system crystallize for LD and MD simulations shown in the panel (a).}
\end{figure}

\begin{figure*}
\includegraphics[width=0.9\linewidth]{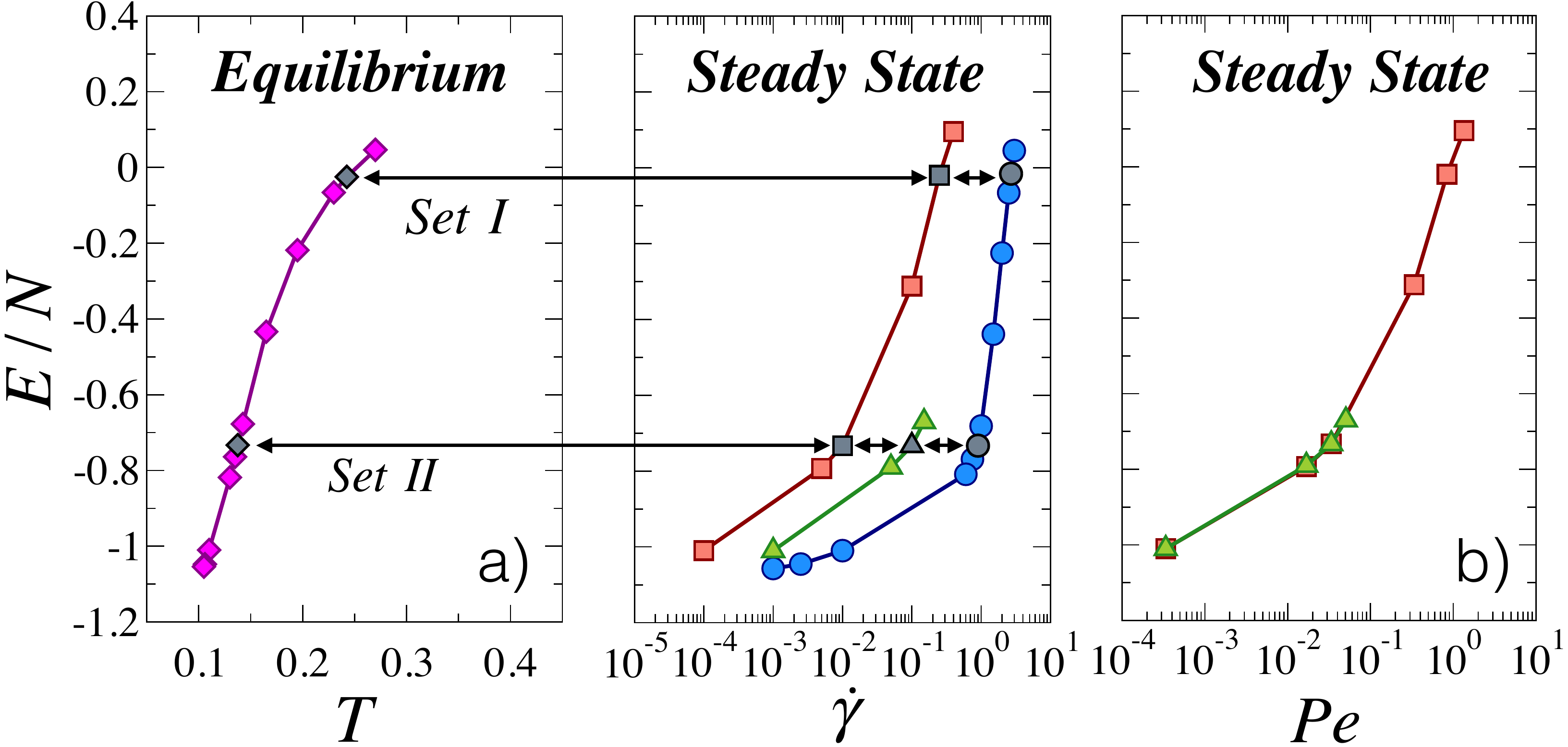}
\caption{\label{fig:Mapping} (a) Mapping $(\dot\gamma, T)$ between equilibrium and steady states via the potential energy per particle $E/N$ in equilibrium and steady states, the latter obtained via MD simulations (circles), LD simulations with $\xi=10$ (triangles) and $\xi=10^2$ (squares). Arrows highlight investigated corresponding states (grey symbols). (b) Potential energy per particle $E/N$ for LD simulations as function of $Pe$.}
\end{figure*}

\subsection{\label{sec:level4bis}Shear-induced crystallization} 

As anticipated in the previous section, we find that our gels undergo crystallization when we perform LD simulations at low enough friction coefficient ($\xi=10$). We have identified the important role of friction which, if too high, counteracts the effect of shear, and acts against ordering. To further strengthen this point, we also perform MD simulations, as described in the Methods, where the solvent is absent, which would thus mimic an atomic, rather than a colloidal, system. 

Crystallization is found when a sudden drop in the potential energy occurs for a given trajectory. To quantify the transition, we monitor the fraction of solid particles $X$ and we calculate the bond local order parameters in order to discriminate between different crystal structures~\cite{lechner2008accurate,russo2012microscopic} (see Methods for details). We report the strain evolution of $X$ for a few selected shear conditions in Fig.~\ref{fig:Crystallization}(a) for both LD and MD simulations. In order to quantify shear also in the case of MD, we refer to the value of the shear rate $\dot\gamma$, because in this case a $Pe$ cannot be defined. We find that crystallization occurs only in a narrow region of  shear rates, which are not too large to be able to destroy the order and not too small in order to induce a significant rearrangement. Thus for the small $Pe$ difference considered in Fig.~\ref{fig:Crystallization}(a), we find that the system sheared with $Pe\approx 2\cdot 10^{-2}$ is able to crystallize, while the system with slightly larger $Pe\approx 3\cdot 10^{-2}$ is not able to crystallize within the simulated time window.
However, an important point is that  the final crystal state is the same for both LD and MD simulations. This is characterized by a predominant face-centered-cubic (fcc) structure, as shown in Fig.~\ref{fig:Crystallization} (b1, b2) where the bond orientational parameter $\bar{w}_{6}$ and $\bar{w}_{4}$ are reported.  A negative value of $\bar{w}_{6}$ can be used to discriminate a fcc arrangement from a body-centered-cubic (bcc) one. However, the fcc structure is quite similar to the hexagonal-close-packed (hcp) one in terms of $\bar{w}_{6}$. In order to discriminate between the latter two crystal structures, one needs to consider $\bar{w}_{4}$, which is predominantly negative for fcc. For even longer simulation times, we find that the system acquires a well-defined fcc crystal at $\gamma\geq10^{4}$. We could only reach this long-time regime using MD simulations within our simulated time window.

\begin{figure}[h]
\includegraphics[width=0.85\linewidth]{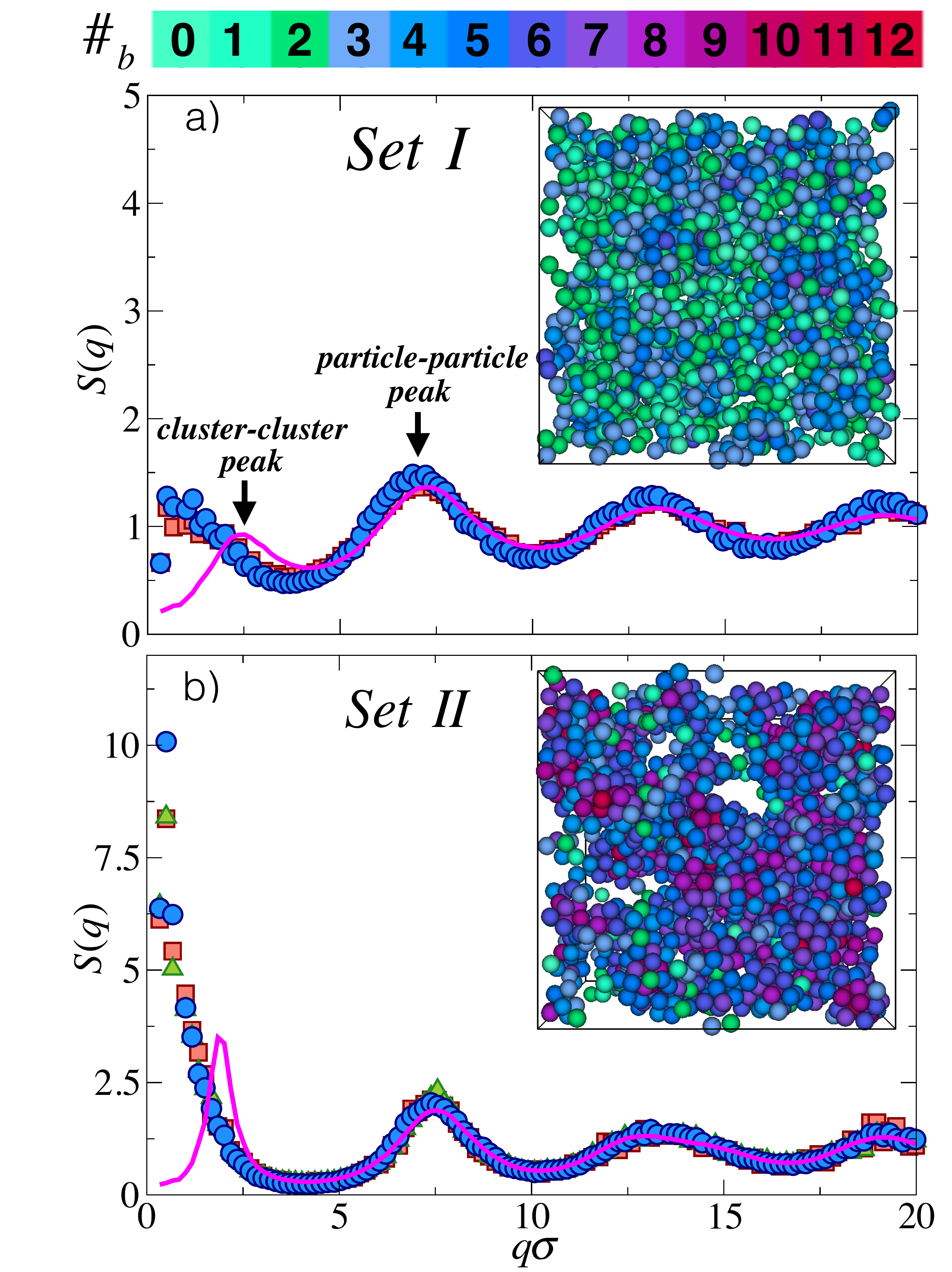}
\caption{\label{fig:Sq_Mapping} Static structure factors $S(q)$ for states with same potential energy and snapshots for MD simulations. Particles are coloured according to their number of bonded neighbours $\#_b$ as indicated in the top colour bar. {\it Set I} (a) and {\it Set II} (b) correspond to the parameters indicated by the mapping shown in Fig.~\ref{fig:Mapping}.}
\end{figure}

\subsection{ Mapping between equilibrium and steady states under shear: structure}
\label{sec:mapping}
Up to this point we have investigated the microscopic structure of the system under shear, quantifying the anisotropy and detecting the onset of crystallization under specific shear conditions. After the yielding point, the system approaches a steady state whose microstructure will be different depending on $Pe$ or on the underlying microscopic dynamics. In particular, we have compared three cases --- the absence of a solvent (MD) and two implicit solvents implemented through LD at respectively low ($\xi=10$) and high viscosity ($\xi=10^2$) --- finding that the solvent can affect the kinetics of the deformation of the initial structure induced by the shear. It would thus be useful to have a way to compare these three cases when shear conditions are equivalent with respect to the underlying Brownian motion (or in the absence of it). To this aim, we can use the equilibrium states of the system as reference states and quantify the effect of the employed shear in each case with respect to them. In particular, we build a correspondence to equilibrium states using the potential energy as mapping observable. Thus, for any applied shear under MD and LD conditions, we consider the potential energy of the steady state and map it to the equilibrium state with the same potential energy. In this way, we establish a shear rate-temperature connection linking steady states obtained under different types of shear and equilibrium states. The obtained mapping is represented in Fig.~\ref{fig:Mapping}(a), where we use $\dot{\gamma}$ as mapping variable in order to include also the MD simulations. 

For each $T$ in equilibrium, a set of corresponding steady states arising from different simulation methods is defined by different values of $\dot{\gamma}$. The weaker is the effect of the solvent, the higher is the shear rate corresponding to the same equilibrium state. To see whether the mapping is meaningful we now consider two sets of states labeled in the Fig.~\ref{fig:Mapping}(a) as {\it Set I} and {\it Set II}, corresponding to states in equilibrium with $T=0.24$ and $T=0.14$ respectively, and we compare the structure of these sets with their corresponding equilibrium states. In Fig.~\ref{fig:Sq_Mapping} we report the static structure factors, calculated as defined in Methods in equilibrium and under shear, for these two sets of corresponding states. Remarkably, we find that the $S(q)$ for all steady states obtained under different shear conditions and dynamics are superimposed onto each other within the statistical uncertainty of the numerical data. This is a confirmation of the efficiency of the mapping in connecting steady states at different shear rates among themselves, and with respect to equilibrium. Thus if one wants to compare different shear conditions, one needs to consider a different $\dot\gamma$ in order to arrive at a similar steady state structure.

In addition, Fig.~\ref{fig:Sq_Mapping} clearly shows the effect of shear on the microscopic structure of the system. While for equilibrium states a cluster peak is observed, extensively discussed in the literature as a generic feature of competing interactions, such peak disappears in the presence of shear in favour of a growing intensity of $S(q)$ for $q\rightarrow 0$. This allows us to deduce that, in the steady state, the shear flow acts essentially at large length scales, by screening the contribution of the long-range repulsion and enhancing the attractive interactions between the colloids as compared to equilibrium case. In this way the shear drives the system closer to phase separation and the presence of larger density fluctuations with respect to equilibrium also helps crystallization of the system. This is visible in the snapshots  reported in Fig.~\ref{fig:Sq_Mapping}(b), where the inhomogeneity of the structure is evident and confirms the findings of enhanced anisotropy after yielding reported in Fig.~\ref{fig:Anisotropy}(d) prior to crystallization. It is important to notice that for steady states corresponding to lower $T$ (not shown), we find no growth of $S(q)$ at small wavevectors and consequently no crystallization. Finally, focusing on wavevectors larger than the nearest-neighbour peak, we see that the shear has a much weaker effect, leaving almost unaltered the local structure of the system. Hence, in our gels with competing interactions, shear essentially acts against the long-range repulsion and is able to strengthen the attractive interactions. It would be interesting to repeat this analysis for other types of gels in order to highlight the different effect of shear in those cases.

\indent 
\begin{figure}[h]
\includegraphics[width=\linewidth]{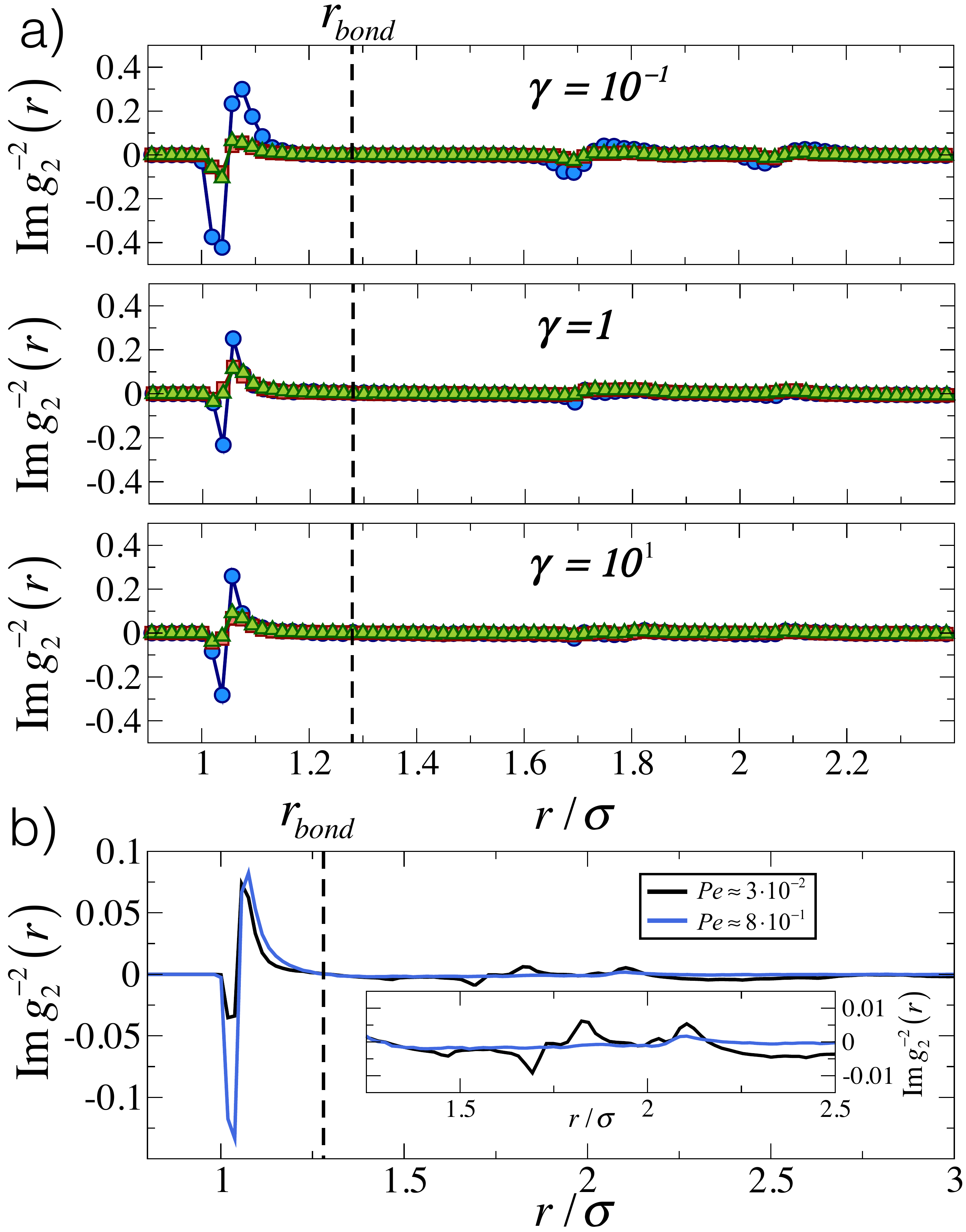}
\caption{\label{fig:An_Mapping} (a) Evolution of anisotropy Im$g_{2}^{-2}\left(r\right)$ at different values of strain $\gamma$ for LD simulations with $\xi=10$ (triangles) and $\xi=10^{2}$ (squares), as well as MD simulations (circles), corresponding to {\it Set II} of Fig.~\ref{fig:Sq_Mapping}. (b) Anisotropy in the steady state for $\xi=10^{2}$ for Set I and Set II state points of Fig.~\ref{fig:Sq_Mapping}. {\it Inset:} Zoom for $r>r_{bond}$ to highlight the presence of anisotropy even at intermediate length scales.}
\end{figure}


\subsection{Comparing different steady states under shear through the established mapping: invariance on P\'eclet number, but dependence on microscopic dynamics}
\label{sec:level5}
The established mapping not only allows us to identify the effect of the shear on the structure of the system, but also allows us to compare the rheological response of corresponding steady states. If the mapping is meaningful, this response should also be identical. We also notice that,  from the definition of the P\'eclet number (Eq.~\ref{eq:Pegel}), there are different possible combinations of $\xi$ and $\dot{\gamma}$ that allow the same $Pe$ value to be obtained. In Fig.~\ref{fig:Mapping}(b)  the potential energy per particle $E/N$ for LD simulations at the two different values of $\xi$ studied here is shown as a function of $Pe$ instead of shear rate. Thus it is evident that our mapping implies that state points with  identical $Pe$ also have the same potential energy, confirming that it is only the balance between shear rate and  Brownian motion that should determine the rheological response of the system. 


We now compare steady states obtained for LD simulations with $Pe\approx3\cdot10^{-2}$ and varying $(\dot{\gamma}, \xi)$ combinations, as highlighted in Fig.~\ref{fig:Mapping}(a) with label {\it Set II}, where the steady states have the same potential energy. In Fig.~\ref{fig:An_Mapping}(a) we show that steady states obtained under shear for the same $Pe$ not only possess an identical structure but also display identical anisotropy distribution at all strain values. We confirm that the same results also hold also for different sets of corresponding states showing asymmetric features of Im$g_{2}^{-2}\left(r\right)$ as in Fig.~\ref{fig:Anisotropy}(d) (not shown).
While this may seem like an obvious result, it offers the possibility to vary independently the two parameters in LD simulations (i.e. $\xi$ and $\dot{\gamma}$) in order to investigate different $Pe$ regimes. So, for example in our system, while using $\xi=10$ and too high values of shear rates, the temperature does not remain constant preventing us to  explore high values of $Pe$ at this effective viscosity. However, building on the mapping, one can equivalently explore higher values of $Pe$ by increasing $\xi$ and using a smaller value of shear rate. Similarly, for example to study shear-induced crystallization may require very long simulation times at high solvent viscosities (and indeed we are not able to detect it within the duration of our simulations). However, a proper balance of the choice of $\dot{\gamma}$ and $\xi$ can be tailored for the specific needs of a given situation, allowing to explore the parameter space in a much more efficient way, without affecting the structure of the final state.


\begin{figure}[h]
\includegraphics[width=\linewidth]{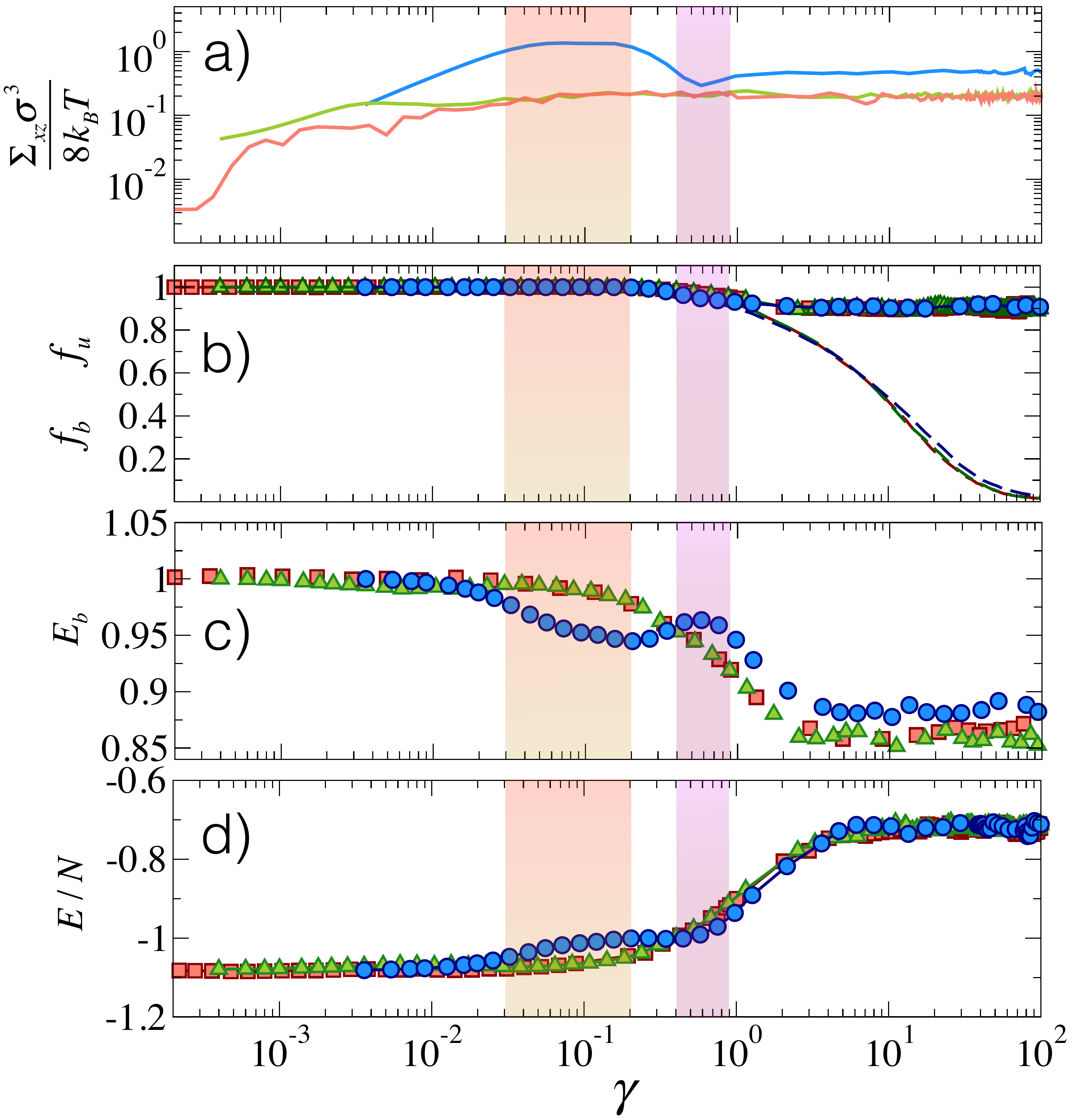}
\caption{\label{fig:MD_LD} Comparison of gel response under shear for LD simulations with $\xi=10$ (green lines/triangles) and $\xi=10^{2}$ (red lines/squares), as well as for MD simulations (blue lines/circles), corresponding to {\it Set II} of Fig.~\ref{fig:Sq_Mapping}: (a) normalized stress versus strain. The two shaded areas highlight the strain regions in which yielding takes place and the MD data show a bump, respectively; (b) fraction of bonds $f_{b}$ (solid lines), fraction of unbroken bonds $f_{u}$ (dashed lines); (c) bond energy $E_{b}$, normalized to their zero-shear values and (d) shows the potential energy per particle $E/N$.}

\end{figure}

We also find that the mapping does not hold  when we also consider MD simulations. Fig.~\ref{fig:An_Mapping}(a) also shows that anisotropy is always higher in the absence of the solvent. This indicates that the presence of the Brownian motion mitigates the growth of the anisotropy and hence, different rheological response are observed when using MD and LD simulations. 

Furthermore, in Fig.~\ref{fig:An_Mapping}(b) we show the anisotropy persisting also in the steady state for $\xi=10^{2}$ for different values of Pe, corresponding to Set I and Set II state points in Fig.~\ref{fig:Sq_Mapping}.
For the larger $Pe$ value, the number of bonds in the steady state is small and the structure of the system is composed of many small groups. On the other hand, for the smaller Pe,  where the system is closer to phase separation, the system is found in a new gel state, which is able to maintain a large amount of anisotropy, which propagates through the structure even well beyond the bond distance.
 
To understand the full rheological response of our gel, we also study the mechanical response versus strain for Set II and again we  focus on the role played by the underlying microscopic dynamics. We plot in Fig.~\ref{fig:MD_LD}(a) the behavior of stress versus strain which clearly shows that the two LD simulations yield an identical behavior at large strains, while a dependence on the effective viscosity $\xi$ is observed for $\gamma \lesssim10^{-2}$, where the effects of the Brownian dynamics balance the shear flow in a different way. On the other hand, the MD simulations display a much larger accumulated stress, with a yielding point that takes place at a different strain value, which is then followed by an oscillation before reaching a steady state with a stress larger than that found for LD simulations. To microscopically compare the three different simulation methods, we again consider the fraction of bonds and the fraction of unbroken bonds versus strain, shown in Fig.~\ref{fig:MD_LD}(b) and we find that these observables do not show a dependence on the shear conditions and on the solvent effects at all strains. However, the energy of the bonds reported in Fig.~\ref{fig:MD_LD}(c) does show significant differences between the MD and LD simulations. In particular, the MD data display an anticipated decrease of $E_b$, associated to their own yielding point, followed by an oscillation which reflects that observed in the stress. The long-time limit of $E_b$ in this case is different from that obtained in LD simulations, despite the potential energy being the same as imposed by our mapping and shown in Fig.~\ref{fig:MD_LD}(d). Thus, we conclude that, as expected in the absence of the solvent, the shear has a much stronger effect on the system: despite having the same potential energy and structure, the spatial configuration of the bonds is rather different as the stress tensor and $E_{b}$ show, indicating that the Brownian motion acts as an additional relaxation mechanism against shear flow. 

\begin{figure}[h]
\includegraphics[width=\linewidth]{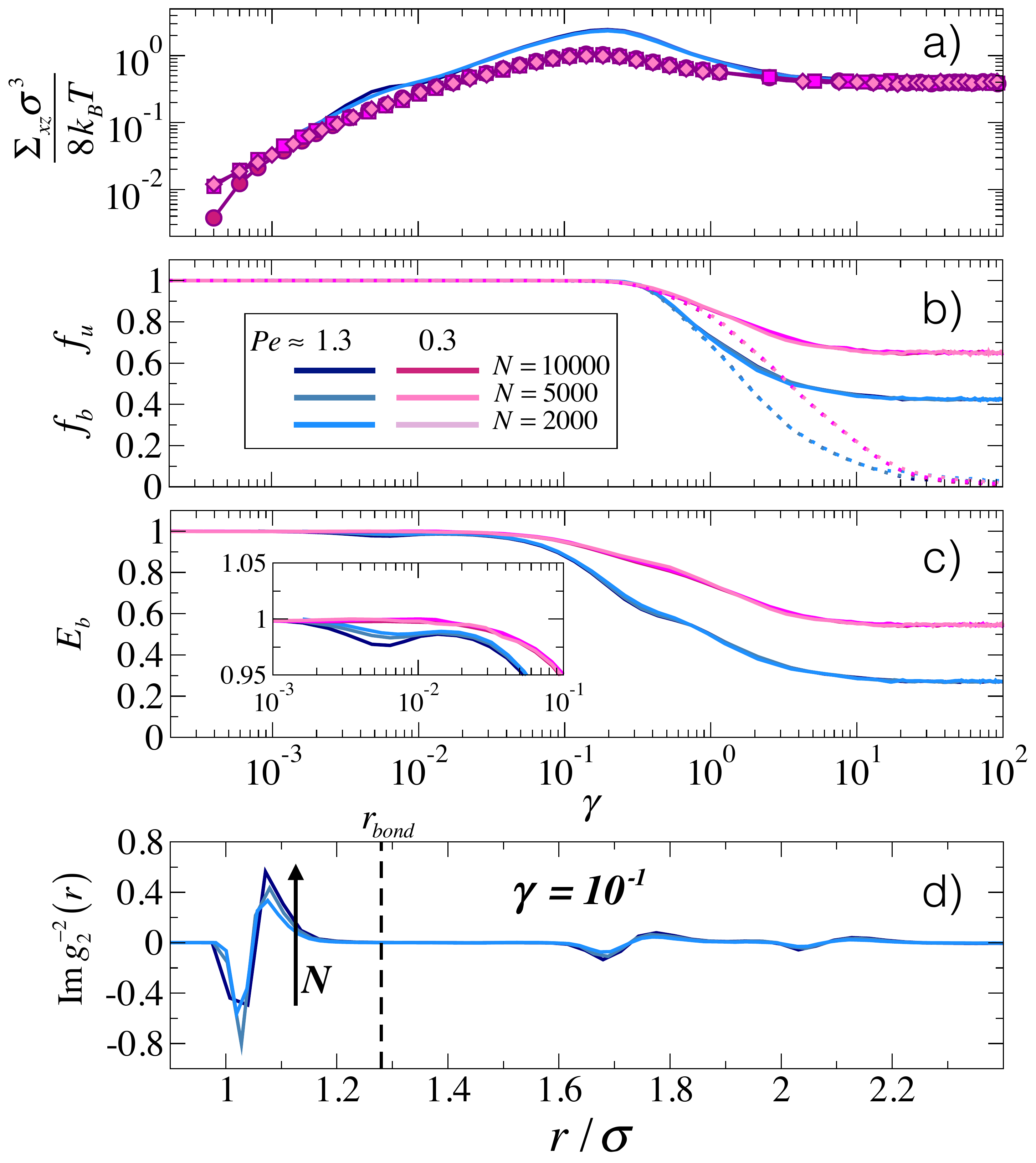}
\caption{\label{fig:SizeEffects} Comparison of gel response under shear for LD simulations with $\xi=10^{2}$ and two values of $Pe$ (as indicated in the legends) at different system sizes $N$: (a) normalized stress versus strain; (b) fraction of bonds $f_{b}$ (solid lines), fraction of unbroken bonds $f_{u}$ (dashed lines) and (c) bond energy $E_{b}$, normalized to their zero-shear values. Inset: zoom of the low strain regime, $10^{-3}\leq\gamma\leq10^{-1}$, to highlight that the bond energy shows oscillations before the yielding point at $Pe>1$; (d) anisotropy Im$g_{2}^{-2}\left(r\right)$ at the yielding point $\gamma=10^{-1}$.}
\end{figure}



\subsection{\label{sec:level7}Size effects}

In this section, we investigate whether and how the size of the system affects the results reported so far. Stress versus strain curves are reported in Fig.~\ref{fig:SizeEffects} for three system sizes at two different values of $Pe$ for LD simulations with $\xi=10^2$. We find that the curves are  superimposed onto each other at all investigated $N$, except for some small differences at short times. Likewise, the quantities $f_{b}$, $f_{u}$ and $E_{b}$ remain unchanged in the stationary state as shown in Fig.~\ref{fig:SizeEffects}(b). Interestingly, in the transient regime, when the velocity profile has not yet matched the imposed one, we find that the bond energy slightly decreases on increasing $N$ (Fig.~\ref{fig:SizeEffects}(c)) while the number of bonds remains constant. This occurs for $Pe>1$ in a small ${\gamma}$ window where shear flow acts on timescales comparable to those of particle diffusion, thus influencing the behavior of the system. In this regime bonds start to stretch as indicated by a decrease of $E_{b}$. However, on increasing the bond distance, particles feel the presence of a repulsive shoulder in their interaction thanks to which $E_{b}$ increases again as shown in the inset of Fig.~\ref{fig:SizeEffects}(c) where it forms a sort of oscillation at small ${\gamma}$. Once the shear flow attains the imposed velocity profile, shear effects occur at a shorter time scale than the relaxation mechanisms and hence, the bonds stretch until they break. Such behaviour is more evident on increasing the system size due to the larger signal coming from bond stretching. The bond energy oscillations disappear for $Pe<1$, indicating that the shear flow do not affect considerably particle dynamics but can be considered as a perturbation of particle Brownian motion. The expression of different relaxation mechanisms in the transient regime has been recently reported in depletion gels~\cite{johnson2018yield}. However, in that case the oscillations in the bond energy were not reported.  Finally we monitor the anisotropy distribution in (Fig.~\ref{fig:SizeEffects}(d)) at the yielding point. We find that there is a small increase of anisotropy with $N$, but overall size effects are not very pronounced and do not qualitatively change the observed patterns. We also note that our results are also in qualitative agreement with simulations performed on much larger system sizes~\cite{moghimi2017residual,johnson2018yield}. Thus we can conclude that the results obtained for $N=2000$ particles are robust and qualitatively representative of larger system sizes for the considered properties, that can be considered ``bulk'' properties. Of course, we may expect some size dependence for the microscopic behavior and this will be addressed in future work.

\section{\label{sec:level8}Discussion and Conclusions}
In this work we have investigated the rheological behavior of gels under steady shear with different numerical approaches. While several experimental and numerical studies have already addressed this problem for depletion-induced gels, which are formed out-of-equilibrium via an arrested spinodal decomposition, we have focused on equilibrium gels, obtained via the competition of depletion attraction and electrostatic repulsion. The main purpose of this work was thus to understand how the response of gels under shear is affected by the route by which the gel is obtained, and hence how it depends on the inter-particle potential between the particles. To reach this goal, we have performed three different types of simulations. Two sets of LD simulations were run for different friction coefficients, tuning in this way the effective viscosity of the implicit solvent. These were then compared to MD simulations where the presence of the solvent is neglected. In this way we could compare conditions which can describe Brownian colloidal motion with others which describe atomic dynamics. While the use of MD simulations in some cases is acceptable also for colloidal systems, for example when focusing on the slow dynamics only\cite{gleim1998does}, under shear the effect of the solvent becomes relevant.  Although LD simulations do not take into account hydrodynamic interactions, they provide a more realistic approach than MD in order to compare with experiments. It is fair to say that our work represents one of the few examples providing a systematic comparison of the influence of microscopic dynamics under steady shear.  

We have calculated the stress tensor for our gels with competing interactions, finding that they exhibit one yielding point prior to reaching a liquid-like steady state (Fig.~\ref{fig:Stress}). The strain at which the yielding point occurs is found to be independent of $Pe$. These results are in agreement with depletion-induced gels at comparable packing fractions and attraction strengths~\cite{koumakis2011two}. Similarly to these studies, the stress overshoot displays a power-law dependence on P\'eclet number, but with an exponent that is higher than values obtained for depletion gels in both experiments and numerical simulations. We attribute this increment to a better resistance of the studied type gel to shear flow thanks to the long-range electrostatic contribution in the colloid-colloid potential. These findings suggest that the rheological response of a colloidal gel can be systematically varied by changing the effective interparticle interactions. Such a feature is very appealing for practical applications and for achieving a fine control of the rheological properties of a gel.

From the microscopic point of view, we have analysed the effect of strain on the gel structure for different $Pe$. In agreement with previous works on different types of gels~\cite{boromand2017structural,johnson2018yield}, we find that the initial bonds forming the gel are deformed after switching on the shear, and start to break only after the yielding point. However, they soon reorganize into a new network structure, whose characteristics depend on the P\'eclet number and also on the microscopic dynamics.

Also similarly to previous works~\cite{park2017structure,johnson2018yield,jamali2017microstructural}, we find that the maximum anisotropy is reached at the yielding point. After this point, the initial gel structure is lost due to the bonds breakage, thereby decreasing the amount of stored stress and consequent anisotropy. At long times or large strains,  the gel approaches a steady state which maintains some degrees of anisotropy at short length-scales~\cite{park2017structure,jamali2017microstructural}. To quantify anisotropy, several works have reported the so-called fabric tensor~\cite{boromand2017structural,johnson2018yield}, even resolved along different directions~\cite{jamali2017microstructural}. Other works instead focus on the expansion of the pair correlation function~\cite{park2017structure,moghimi2017residual}, including the present work. These different observables provide similar amount of information on the anisotropy, but the use of Im$g_{2}^{-2}\left(r\right)$ allows one also to obtain spatial resolution. In general,  Im$g_{2}^{-2}\left(r\right)$ is found to have two roughly symmetric peaks, localized around the bond distance and of roughly maximum intensity at the yielding point. The balance between compressional (negative) and extensional  (positive) peak depends on the  employed $Pe$ as well as on the specific potential interaction. In our system, this holds for $Pe>10^{-1}$. For certain conditions, an asymmetric situation is found, where a large positive peak is accompanied by an almost absent negative peak. We find that this situation occurs for $Pe<10^{-2}$ and we suggest that this feature is a precursor of a fluid-to-crystal transition, which is obtained only in a narrow region of $Pe$ and at low enough solvent friction (at least within our simulation time window). Thus, the accumulation of particles in the extensional axis Im$g_{2}^{-2}\left(r\right)$ seems to be a pre-requisite for a transition from an amorphous to an ordered structure. Conversely, in those cases where the asymmetry is not found, crystallization is hampered because either the small crystal nuclei are not able to support the deformation induced by the shear flow or the used shear rate values are not able to dominate over the underlying Brownian dynamics. Interestingly, in Ref.~\cite{moghimi2017residual}, an increase of the extensional peak is also found for low $Pe$. Differently, from previous works on gels, our inter-particle potential includes a long-range repulsion, which is found to affect the anisotropy at the yielding point even at  large length scales well beyond the bond distance. 

Finally we also note that, while in general at low strains the anisotropy only acts at the direct level of bonds, at the yielding point it also affects large length scales well beyond the bond distance. The fact that the anisotropy is clearly observed also for $r>r_{bond}$ seems to be a distinctive feature of the present work, where the presence of an additional long-range Yukawa repulsion is responsible for the propagation of the anisotropy at larger scales, differently for depletion-like gels\cite{park2017structure,moghimi2017residual} but in agreement with numerical simulations of repulsive Yukawa glasses\cite{zausch2009build}.

To be able to connect results obtained with different types of simulations, we built on ideas borrowed from studies of the glass transition where states in equilibrium and in aging are connected through a time-temperature relation. In a similar fashion, we consider here a mapping between equilibrium and steady states under shear (so-called corresponding states) by considering state points at different $\dot\gamma$ and $T$ with the same potential energy (Fig.~\ref{fig:Mapping}). This mapping confirms the intuitive expectation that, upon decreasing the effects of the solvent, higher values of shear rates are required in order to reach a similar steady state. We find that the static structure factors of the examined corresponding steady states are identical to each other (Fig.~\ref{fig:Sq_Mapping}). In particular, the cluster peak observed in equilibrium, which is a distinctive feature of gels obtained by competing interactions~\cite{stradner2004equilibrium}, is destroyed by the presence of the shear in favour of a growth of $S\left(q\right)$ for $q\rightarrow 0$. This implies that the shear flow acts at large length scales, screening out the contribution of the long-range repulsion and hence, pushing the system closer to phase separation. Interestingly, these results are in agreement with those obtained by recent simulations of depletion-induced gels in the presence of hydrodynamic interactions~\cite{jamali2017microstructural}. In addition, the mapping puts forward the evidence that the response of the gel and the approach of the steady state is controlled solely by $Pe$ in LD simulations, providing identical results upon varying solvent conditions and shear rates. This confirms that the competition between Brownian motion and shear flow is the key parameter to control the behavior of the gel under shear. However, when we compare equal potential energy states under shear in the absence of the solvent, we find a stronger effect of the shear on the gel in terms of anisotropy and final steady state properties, signaling that the mapping does depend on the choice of the microscopic dynamics. These results can be useful in future simulation works because they allow an independent choice of solvent viscosity and shear rate respectively, maintaining the same $Pe$. This makes it possible to explore a wide range of shear conditions without affecting the resulting steady states. This is particularly valuable from the practical point to view to extend investigations at high $Pe$ for example ensuring temperature stability as well as to explore low $Pe$ where competing mechanisms inducing by shear may favour crystallization, as in the present case, or other underlying processes. It will be interesting to extend the mapping to different types of simulations where solvent is treated in a more accurate ways including hydrodynamics effects, e.g. using Dissipative Particle Dynamics \cite{zausch2008equilibrium,ruiz2018crystal} or Multi-Particle Collision Dynamics\cite{ripoll2005dynamic,kapral2008multiparticle}.



Finally, we have shown that our results are robust against the variation of the system size. Thus, the modification of the inter-particle interaction is able to provide a different rheological response with respect to the widely studied depletion gels. Our results call for experimental investigations probing the stress-strain behavior of gels with competing interactions, still missing so far. In addition, it would be interesting in the future to extend our study to different types of gels, particularly equilibrium gels resulting for patchy or limited-valence attractions\cite{sciortino2017equilibrium} for which rheological investigations are scarce.

\begin{acknowledgments}
JRF, EZ acknowledge support from ETN-COLLDENSE (H2020-MCSA-ITN-2014, Grant No. 642774). NG, EZ from ERC Consolidator Grant 681597 MIMIC.  We thank Daniele Parisi for useful discussions.
\end{acknowledgments}

\bibliography{biblio}

\end{document}